\documentclass[aps, prl, twocolumn, preprintnumbers, nofootinbib, superscriptaddress, balancelastpage]{revtex4-2}

\usepackage[utf8]{inputenc}
\usepackage{amsmath, amsfonts, amssymb}
\usepackage{bm,bbm}
\usepackage{hyperref}
\usepackage{xcolor}
\usepackage[dvipsnames]{xcolor}
\usepackage{comment}
 \usepackage{tikz}
\newcommand{\beq}{\begin{eqnarray}}
\newcommand{\eeq}{\end{eqnarray}}
\newcommand{\beqa}{\begin{equation}\begin{aligned}}
\newcommand{\eeqa}{\end{aligned}\end{equation}}

\usepackage{scalerel} 
\newcommand{\prlsec}[1]{{\bf\emph{#1:}}}

\newcommand{\scym}{\scaleto{\rm YM}{0.8ex}}
\newcommand{\bw}{\scaleto{\rm BW}{0.8ex}}
\usepackage{subcaption}
\begin{abstract}
The thermal confinement phase transition (PT) in $SU(N)$ Yang-Mills theory is first-order for $N\geq 3$, with bounce action scaling as $N^2$. Remarkably, lattice data for the action include a small coefficient whose presence likely strongly alters the PT dynamics. We give evidence, utilizing insights from softly-broken SUSY YM models, that the small coefficient originates from a deconfined phase instability just below the critical temperature.  We predict the maximum achievable supercooling in $SU(N)$ theories to be a few percent, which can be tested on the lattice. We briefly discuss the potentially significant suppression of the associated cosmological gravitational wave signals.
\end{abstract}

\begin{document}

\title{Prediction for Maximum Supercooling in SU(N) Confinement Transition}
\author{Prateek Agrawal}
\email{prateek.agrawal@physics.ox.ac.uk}
\affiliation{Rudolf Peierls Centre for Theoretical Physics, 
University of Oxford, Parks Road, Oxford OX1 3PU, United Kingdom}
\affiliation{Department of Physics, University of California, Santa Barbara, CA 93106, USA}
\author{Gaurang Ramakant Kane}
\email{gaurang.kane@physics.ox.ac.uk}
\affiliation{Rudolf Peierls Centre for Theoretical Physics, 
University of Oxford, Parks Road, Oxford OX1 3PU, United Kingdom}
\author{Vazha Loladze}
\email{vazha.loladze@physics.ox.ac.uk}
\author{John March-Russell}
\email{john.march-russell@physics.ox.ac.uk}
\affiliation{Rudolf Peierls Centre for Theoretical Physics, 
   University of Oxford, Parks Road, Oxford OX1 3PU, United Kingdom}

\maketitle

\prlsec{Introduction} 
The thermal phase transition (PT) between the deconfined and confined phases of $SU(N)$ Yang-Mills (YM)  theory is an important open problem in theoretical physics. Consequently, much effort has gone towards understanding this PT using lattice gauge theory, holographic models and other modified versions of the theory with greater control, e.g.~due to supersymmetry. These approaches indicate that the PT is first-order for $N\geq 3$, with the rate scaling as $\exp(-N^2)$ in the large-$N$ limit. Theoretical studies typically focus on the equilibrium thermal state, for which the PT formally occurs at the critical temperature, $T_{\rm cr}$, where the deconfined and the confined phases have equal free energies. 

However, in a dynamical setting with varying temperature, such as a cosmological PT where the universe goes from a deconfined to a confined phase, the PT takes place at a strictly lower temperature. In this regime the PT is governed by the transition from the metastable deconfined phase. It is thus crucial to understand the physics of the metastable phase as the universe supercools below the critical temperature.

This issue of confining PT dynamics is particularly pressing as gravitational waves (GWs) from first-order cosmological PTs provide a new opportunity to investigate physics beyond the Standard Model~\cite{Kosowsky:1992rz,Huber:2008hg,Caprini:2024hue, Reardon:2023gzh,Caprini:2019egz, Athron:2023xlk}. 
Gauge theories with confinement scales around the electroweak scale are especially compelling as candidates for new physics related to the electroweak hierarchy problem~\cite{Kaplan:1983fs,Georgi:1984af,Kaplan:1991dc,Contino:2003ve,Agashe:2004rs,Arkani-Hamed:2001nha,Chacko:2005pe} and to dark matter (see \cite{Kribs:2016cew} and references within). The confinement transition in these theories produces GWs with a frequency peak in the sensitivity range of upcoming experiments such as LISA~\cite{Grojean:2006bp}.
The supercooling parameter $\epsilon \equiv 1-T/T_{\rm cr}$ plays a crucial role in the phenomenology of cosmological PTs and their associated GW signatures. 

A first-order PT is a necessary condition for generating a sizeable GW signal from the PT bubble collisions.  A strong signal, however, additionally requires both strong supercooling and far-from-quasi-equilibrium dynamics of the bubble expansion. For small supercooling, the PT timescale parameter $\beta/H \sim 1/\epsilon^{3}$ is larger, suppressing the strength of the GW signal. Small $\epsilon$ also suppresses the bubble wall velocity. Overall, even if the dynamics of the PT is qualitatively unaltered, small supercooling can suppress the GW signal by a few orders of magnitude.

However, at small supercooling there can be much more dramatic effects on the subsequent bubble dynamics and the resultant GW signal.  As each bubble expands, the latent heat of the PT reheats the plasma surrounding the bubble. The bounce action near the critical temperature scales rapidly as $\epsilon^{-2}$, and the rate of bubble nucleation grows exponentially. This can result in an efficient reheating of the universe to close to the critical temperature, such that the pressure differential across the bubble wall becomes small, and the walls expand with negligible speed~\cite{GarciaGarcia:2015fol}. In such a case the gravitational wave production is suppressed to insignificant levels. 

In the case of confining PTs, the inherently strong coupling complicates the detailed calculation of the PT dynamics.
Determining the exact range of supercooling where this effect can be observed depends on the details of plasma properties such as the heat transport coefficient and the specific heat of the supercooled plasma~\cite{GarciaGarcia:2015fol}, which is outside the scope of the current work. A simplified simulation was carried out for $SU(3)$ by~\cite{Asadi:2021pwo, Gouttenoire:2023roe} with a transition around the electroweak scale and it was found that the bubble wall velocity $v_{\rm w}\lesssim 10^{-6}$ in that case. Thus small supercooling can completely modify the dynamics of the PT and the associated GW signal.

In this letter we argue that the metastable deconfined phase in $SU(N)$ YM theory becomes unstable just below the critical temperature and provide an estimate for maximum supercooling in the confinement PT. We derive our estimate by combining results from lattice data for $SU(N)$ YM theory and analytical results from an $\mathcal{N}=1$ supersymmetric $SU(N)$ gauge theory with softly broken supersymmetry, which has a calculable analogue of the confinement transition.

\prlsec{Hints from Lattice Studies} We review the lattice results for confining transitions in $SU(N)$ YM~\cite{Lucini:2005vg, Salami:2025iqq}. It is well-established that the PT with $N\geq 3$ colours is first order \cite{Panero:2009tv}, and therefore proceeds via bubble nucleation. To estimate the PT rate near $T_{\rm cr}$, we can apply the $O(3)$-symmetric thin-wall approximation and derive the Boltzmann suppression $\exp(-S_{\rm b})$ with
\begin{eqnarray}
    S_{\rm b}=\frac{16\pi}{3}\frac{\sigma_{\bw}^3}{\Delta f^2 T_{\rm cr}}  \,.
\end{eqnarray}
Here $\sigma_{\bw}$ is the bubble wall tension between the true (confined) and false (deconfined) vacuum, and $\Delta f$ is the temperature-dependent free energy density difference between the deconfined and confined phases with $\Delta f(T_{\rm cr})=0$. We can write $\Delta f$ up to first order in $\epsilon$ as $\Delta f(T)=Q_{\rm h}\epsilon$, where $Q_{\rm h}$ is the latent heat of the PT. Finally, for $\epsilon \ll 1$ we have
\begin{eqnarray}
    S_{\rm b}\approx\frac{16\pi}{3}\frac{\sigma_{\bw}^3}{Q_{\rm h}^2T_{\rm cr}}\frac{1}{\epsilon^2} \,.
    \label{eqn:thin_wall_o3_action}
\end{eqnarray}

From large-$N$ counting, the expectation is $S_{\rm b}\approx O(1) (N^2/\epsilon^2)$. This would result in a highly suppressed nucleation rate and thus a strongly supercooled PT. However, this estimate is significantly mistaken (see~\cite{GarciaGarcia:2015fol} for a discussion at $N=3$). The quantities $Q_{\rm h}$ and $\sigma_{\bw}$ can be extracted from lattice studies~\cite{Lucini:2005vg, Salami:2025iqq}: 
\begin{align}
    Q_{\rm h}
    &=
    \left(0.7731(12)-\frac{0.959(58)}{N^2}\right)^4 N^2T_{\rm cr}^4 \,,
    \\
    \sigma_{\bw}
    &=
    \left(0.0189(11) - \frac{0.190(19)}{N^2}\right) N^2 T_{\rm cr}^3
    \,.
    \label{eq:latticeresults}
\end{align}
Using these results we have in large-$N$ limit, and in the immediate vicinity of $T_{\rm cr}$,
\begin{eqnarray}
S_{\rm b}\approx 8.86\times10^{-4}\frac{N^2}{\epsilon^2}\left(1-\frac{10.05}{N^2}\right)^3 \,.
\label{eqn:o_3_ym}
\end{eqnarray}
This is much smaller than our initial estimate by a factor of $\sim 10^{-3}$, which is a result of the unexpectedly small value of the domain wall tension in Eq.~\eqref{eq:latticeresults}. This suppression in $S_{\rm b}$ is very surprising because the theory seems to contain no small / large parameters except $N$, which we have already taken into account.

Our central point is that this suppression not only significantly reduces the bounce action near $T_{\rm cr}$ but hints at a much more drastic consequence: the metastable deconfined phase exists only for temperatures $T>T_{\rm min}$ with $(T_{\rm cr}-T_{\rm min})/T_{\rm cr}\ll 1$. This explains the smallness of the domain wall tension at $T\sim T_{\rm cr}$ because it vanishes at $T_{\rm min}$ very close to $T_{\rm cr}$ (and as we will argue the functional form of the $T$-dependence for $T_{\rm min}<T<T_{\rm cr}$ is strongly constrained). The small coefficient in Eq.~\eqref{eqn:o_3_ym} is directly correlated with small supercooling in YM confinement transitions.

It is interesting to compare this consequence in various controlled examples that are analogous to the confinement transition.
In holographic setups~\cite{Buchel:2009bh, Morgante:2022zvc, Mishra:2023kiu,Mishra:2024ehr}, it is generic to find a minimal temperature $T_{\rm min}$ close to $T_{\rm cr}$. The gravitational dual to the gauge theory deconfined phase does not exist below $T_{\rm min}$. The value of $T_{\rm min}/T_{\rm cr}$ is model-dependent, but a common feature is that these two temperatures are close to each other hence leading to small supercooling ($\sim 5-10\%$). We will see that this expectation is reproduced in the analogue model with softly broken $\mathcal{N}=1$ supersymmetry.

\prlsec{Softly Broken SUSY on $\mathbb{R}^3\times S^1$} We consider the $\mathcal{N}=1$ SYM theory with a gauge group $SU(N)$ on the manifold $\mathbb{R}^3\times S^1$, with periodic boundary conditions for the bosons \emph{and} fermions in the $S^1$ direction. These boundary conditions preserve supersymmetry. We break supersymmetry softly by introducing a small gaugino mass $m$. The approximate supersymmetry in this case gives us an analytical handle on the theory.

There is a phase transition analogous to confinement in this theory as a function of $L$, the length of the $S^{1}$ direction.
When $m$ is much smaller than the dynamical scale of the gauge theory,  $\Lambda$ (defined as the confinement scale when the theory is placed on Minkowski spacetime) this PT occurs in the weak coupling regime and therefore is calculable analytically. In the limit $m\gg \Lambda,  L^{-1}$, the gauginos decouple before the PT, leaving us with a YM theory and the PT becomes a thermal PT with the $S^1$ identified as a thermal circle with temperature $T=1/L$. However, in this limit, the PT is no longer at weak coupling so analytical control is lost. 

If, following Refs.\cite{Poppitz:2012sw, Poppitz:2012nz, Anber:2013sga, Anber:2017tug}, we assume that the regimes $m\ll \Lambda$ to $m\gg \Lambda$ are smoothly connected, we can derive interesting consequences for the thermal confinement/deconfinement PT in YM theory from the analytically calculable (non-thermal) PT in softly broken SUSY. Thus we now review results related to confinement in the model (see~\cite{Poppitz:2012sw, Poppitz:2012nz, Anber:2013sga, Anber:2017tug} for more details). 

It is well-known that the 1-form $\mathbb{Z}_N^{(1)}$ global symmetry of the Yang-Mills theory descends to a 0-form global symmetry (referred to as the center symmetry) and a 1-form symmetry when we compactify the theory on the thermal circle. The Polyakov loop
\begin{equation}
    \Omega(\vec{x})
    =
    \frac{1}{N}
    {\rm tr \,} \mathcal{P}\left(\exp\left\{i\int_{0}^{L} dx^{4}A^{a}_{4}(\vec{x},x^{4})T^{a}\right\}\right)~~
    \label{eqn:general_poyakov_term}
\end{equation}
transforms under center symmetry as $\Omega\rightarrow e^{\frac{2\pi i}{N}}\Omega$.
The Polyakov loop plays the role of the order parameter in the thermal confinement/deconfinement transition.
If $\langle \Omega\rangle\neq 0$ then the center symmetry is spontaneously broken, corresponding to the deconfined phase of the theory. In the confined phase $\langle\Omega\rangle= 0$ and the symmetry is restored. Therefore to study the PT one needs to obtain the effective potential for the Polyakov loop. 
\begin{figure*}[t!]
    \centering
    \includegraphics[width=0.48\linewidth]
{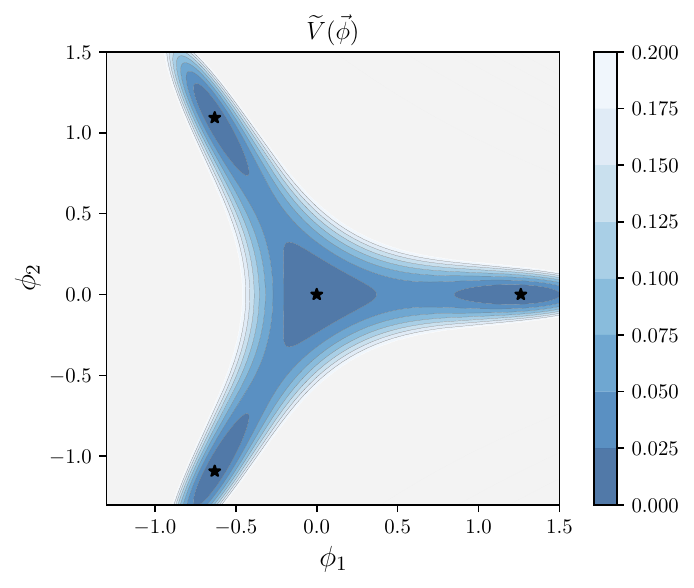}
\includegraphics[width=0.48\linewidth,height=0.38\linewidth]{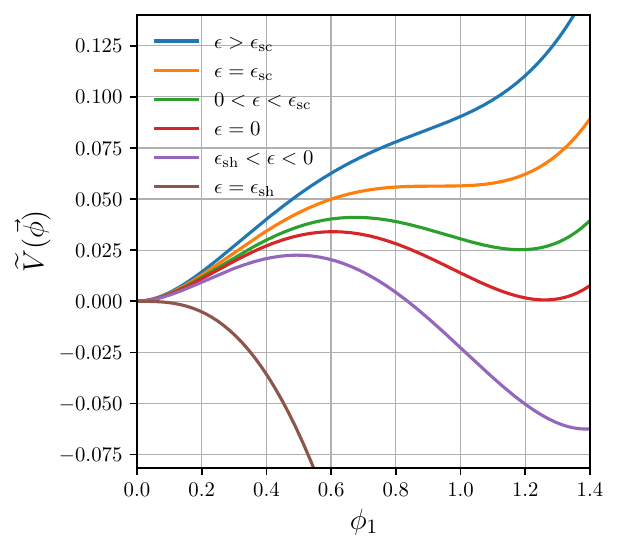}
    \caption{\textbf{Left}: Potential of Eq. \eqref{eqn:approx_pot} with $N=3$ as a function of $\vec{\phi}=(\phi_{1},\phi_{2})$ at $\epsilon=0$. The simplicity of this particular $N=3$ potential is misleading. Although the $\mathbb{Z}_{N}$ structure is present for $N>3$, the valleys connecting the vacua are no longer on straight paths in the field space. \textbf{Right}: Potential for $N=3$ as a function of $\vec{\phi}=(\phi_{1},0)$ for different values of $\epsilon$. For $\epsilon>\epsilon_{\rm sc}$ the deconfined phase is mechanically unstable.}
    \label{fig:su(3)_2d_pot}
\end{figure*}
One can choose a gauge where the only non-zero components of $A^a_{4}$ lie in the Cartan subalgebra of $SU(N)$. We will integrate out massive KK modes and keep only zero modes of $\vec{A}_{4}$ (the vector denotes the Cartan vector space). We will see that the dynamics picks a vacuum close to the centre-symmetric point, so we expand $\vec{A}_4$ as,
\begin{equation}
   \vec{A}_{4}=\dfrac{2\pi}{N L}\vec{\rho}+\dfrac{g_{4}^{2}}{4\pi L}\vec{\phi}
   \,,
    \label{eqn:a_4_decomposition}
\end{equation}
where $\vec{\rho}$ is the Weyl vector. Then, the Polyakov loop is ($\vec{H}$ are the Cartan generators)~\cite{Poppitz:2012nz},
\begin{equation}
    \Omega(\vec{x})=
    \frac{1}{N}
    {\rm tr\,}
    \exp\left\{
    i\dfrac{2\pi}{N} \vec{H}\cdot\vec{\rho}
    +i\dfrac{g_{4}^{2}}
    {4\pi }\vec{H}\cdot \vec{\phi}(\vec{x})
    \right\}\,.
\end{equation}
In this parametrization $\langle\vec{\phi}\rangle=0$ corresponds to the centre-symmetric point, $\langle \Omega\rangle=0$. 

In the 3d effective theory, the vev for $A_4^a$ breaks the $SU(N)$ gauge symmetry down to the Cartan subgroup. Therefore, below the scale of the breaking the light degrees of freedom are the $U(1)^{N-1}$ gauge bosons and their partner fermions, in addition to the holonomies parametrized by $\vec{\phi}$.
There are non-perturbative effects in 3d which generate a gap in the $U(1)^{N-1}$ sector, and provide a potential for $\vec{\phi}$.

The transition in this theory happens in the weak coupling regime, so these effects are calculable. The leading contributions to the effective potential come from: (a) $N$ species of monopole-instantons (of which $N-1$ are monopole-instantons corresponding to each $U(1)$ in the Cartan and one KK monopole-instanton), (b) $N$ species of magnetic bions and (c) $N-1$ species of neutral bions. The latter two are bound states of monopole-instantons (for details see \cite{Poppitz:2012nz}).

We will work with a vanishing $\theta$ angle, and consider only the leading contribution to the potential in the weak coupling, so perturbative contributions to the potential are neglected. In this approximation the potential is,
\begin{equation}
    V (\vec{\phi})
    = V_{0} 
    \sum_{i=1}^{N}e^{-\vec{\alpha_{i}}\cdot\vec{\phi}}
    \left[
    e^{-\vec{\alpha_{i}}\cdot\vec{\phi}}
    -e^{-\vec{\alpha}_{i+1}\cdot\vec{\phi}}
    - \kappa
    \frac{4\pi^2}{N^2}
    \frac{L_{\rm cr}^2}{L^2}
    \right]\,,
    \label{eqn:approx_pot}
\end{equation}
with,
\begin{equation}
   V_{0}=\dfrac{27}{8\pi}\dfrac{\Lambda^{6}}{v^{3}}\ln\left(\dfrac{v}{\Lambda}\right),
   \qquad
   \kappa
    =
    \dfrac{4m}{3\Lambda^3 L_{\rm cr}^2}
    \,.
\end{equation}
The potential depends explicitly on $L$, which is interpreted as $T^{-1}$, and $L_{\rm cr}$ corresponds to the critical temperature where the confined ($\vec{\phi} = 0$) and the deconfined minima are equipotential. 
In the expression above $m$ is the gaugino mass, $v=2\pi/NL$ is the Higgsing scale, $\kappa$ is an $O(1)$ constant, and $\vec{\alpha}_{i}$ are the extended simple roots of the $\mathfrak{su}(N)$ Lie algebra (with length of the root equal to $\sqrt{2}$)~\cite{Georgi:1999wka}. We emphasize that the \emph{structure} of the potential in equation \eqref{eqn:approx_pot} is valid for all $N\geq 2$ with only the number of holonomies $\vec{\phi}$ and detailed numerical coefficients changing.

An important and novel feature of our analysis, which works with a controlled setup, is that the degrees of freedom determining the phase transition and bounce dynamics is not just a single Polyakov loop field variable $\Omega(\vec{x})$ as has been assumed in previous studies \cite{Pisarski:2000eq, Pisarski:2001pe, Pisarski:2002ji, Huang:2020crf}. Instead, our analysis shows that the true variables that determine the transition dynamics are the $N-1$ gauge holonomies $\vec{\phi}(x)$. For the special case of $N=3$, this turns out not to make a significant difference, however, for $N> 3$, the bounce trajectory (that will be commented on later and also discussed with an example in the supplementary material \cite{supp}) is highly non-trivial in the $\phi-$space becoming more so as $N$ increases.

As noted in \cite{Poppitz:2012nz}, for $N\geq 3$, there are some key features of the potential in Eq.\eqref{eqn:approx_pot}. There are two local minima for a range of temperatures $\epsilon_{\rm sh} < \epsilon < \epsilon_{\rm sc}$.
\begin{enumerate}
    \item For $\epsilon_{\rm sc}<\epsilon$, the only minimum is at $\vec{\phi}=0$ corresponding to the center-symmetric \emph{confined} minimum.
    \item For $0<\epsilon<\epsilon_{\rm sc}$ the confined minimum is the global minimum while there exist a set of $N$ local minima related by $\mathbb{Z}_{N}$-symmetry where $\vec{\phi}\neq 0$. They represent the \emph{deconfined} minima where the center symmetry is spontaneously broken (figure \ref{fig:su(3)_2d_pot}).
    \item For $\epsilon_{\rm sh}<\epsilon<0$ the confined minimum is a local minimum and the $N$ deconfined minima are the global minima. 
    \item For $\epsilon<\epsilon_{\rm sh}$ there is no minimum at $\vec{\phi}=0$ and the only minima that exist are the deconfined minima.
\end{enumerate}
We call $\epsilon_{\rm sc}$ and $\epsilon_{\rm sh}$ as the points of maximum possible supercooling and superheating respectively. The above described behaviour can be seen in figure~\ref{fig:su(3)_2d_pot} explicitly for $SU(3)$. (Note that while we argue for the existence and approximate values of $\epsilon_{\rm sc,sh}$ can be reliably established, the actual PT dynamics at
and close to $\epsilon = \epsilon_{\rm sc,sh}$ requires renormalization group resummation due to the existence of IR divergencies arising from thermal fluctuations. \cite{Pablo:2022})

\begin{figure}
    \centering
    \includegraphics[width=\linewidth]{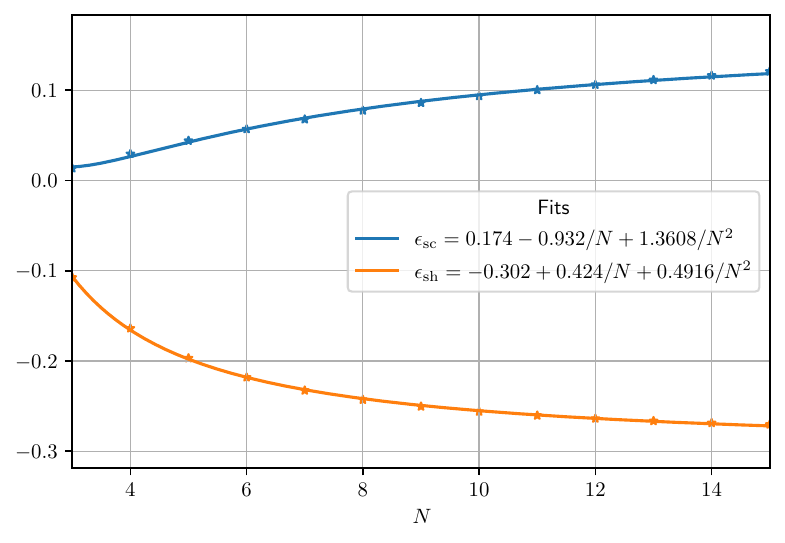}
    \caption{Maximum possible supercooling $\epsilon_{\rm sc}$ and superheating $\epsilon_{\rm sh}$, in the  softly broken $\mathcal{N}=1$ SYM on $\mathbb{R}^3\times S^1$, fitted as a function of $N$ up to order $1/N^{2}$~.}
    \label{fig:maximal_supercooling_superheating}
\end{figure}

Importantly, the values of $\epsilon_{\rm sc}$ and $\epsilon_{\rm sh}$ for different $N$ can be calculated from the potential in Eq.~\eqref{eqn:approx_pot}, with values plotted in figure~\ref{fig:maximal_supercooling_superheating}. It can be seen that both of these quantities have a well-defined large-$N$ limit, although the supercooling parameter converges somewhat more slowly. We can extract further valuable information from this system by numerically calculating the bounce action between the confined and deconfined phases as a function of $\epsilon$. Casting it into an appropriate form amenable to translation to the YM case, we then get some striking predictions for the YM confinement transition.

\prlsec{Calculation of the bounce action} The action for the transition between the two minima takes the form:
\begin{equation}
    S_{\rm b}=\int d^{3}x\left(\dfrac{1}{2}f_{\phi}^{2}(\partial\vec{\phi})^{2}+V(\vec{\phi})\right)
    \label{eqn:bounce_action},
\end{equation}
where the constant $f_{\phi}^{2}$ in front of the kinetic term can be expressed as
\begin{equation}
    f_{\phi}^{2}=\dfrac{g_{4}^{2}}{8\pi^{2} L}=\dfrac{1}{6\pi}\dfrac{v}{\ln\left(\frac{v}{\Lambda}\right)}~~.
\end{equation}
The second equality comes from the two-loop definition of the dynamical scale $\Lambda$. One can factor out the $V_{0}$ from the integral \eqref{eqn:bounce_action} and further rescale the $\mathbb{R}^{3}$ coordinates to scale out all dimensionful quantities, which gives,
\begin{align}
     S_{\rm b}
     &=
     \frac{(v/\Lambda)^3}{27\pi \log^2(v/\Lambda)}
     \int d^{3}y\left(\dfrac{1}{2}(\partial_{y}\vec{\phi})^{2}+\widetilde{V}(\vec{\phi})\right)
     \,,
\label{eqn:factorised_action_r3_s1}
\end{align}
where $\widetilde{V}(\vec{\phi})=V(\vec{\phi})/V_0$. We can express the prefactor in Eq.\eqref{eqn:factorised_action_r3_s1} in terms of the length of the compact direction $L$ and the string-tension $\sigma_{\rm str}$ of the flux tubes generated between the two well-separated Polyakov loops to facilitate translation to YM. 
Following the calculations of \cite{Anber:2017tug}:
\begin{equation}
    \sigma_{\rm str}=\dfrac{9}{2}L\Lambda^3\log\left(\dfrac{v}{\Lambda}\right)~~.
\end{equation}
Note that $\sigma_{\rm str}$ is the far IR value of the string tension in the confined phase (or equivalently $c=0$ in \cite{Anber:2017tug}). Hence, we can write the bounce action as
\begin{align}
  S_{\rm b}
  &=
  \dfrac{N^2\Lambda^3}{\sigma_{\rm str}^2L}\widetilde{S}_{\rm b}~,
  \nonumber\\
  \widetilde{S}_{\rm b}
  &=
  \dfrac{6\pi^{2}}{N^{5}}\int d^{3}y\left(\dfrac{1}{2}(\partial_{y}\vec{\phi})^{2}+\widetilde{V}(\vec{\phi})\right).
    \label{eqn:factorised_action_2}
\end{align}
We study $\widetilde{S}_{\rm b}$ as a function of temperature, or equivalently $\epsilon$. Near $\epsilon=0$, the bounce can be calculated in the 
thin-wall limit, taking the form  given in Eq.~\eqref{eqn:thin_wall_o3_action} which has a double pole at $\epsilon=0$.
Further, we have shown that the action has zeroes at $\epsilon_{\rm sc}$ and $\epsilon_{\rm sh}$ where the barrier between the false and the true vacua vanishes (e.g.~figure~\ref{fig:su(3)_2d_pot}). The large-$N$ scaling of the action also suggests that $\widetilde{S}_b$ should be independent of $N$ in the large-$N$ limit. This motivates the following parametrization for $\widetilde{S}_{\rm b}$:
\begin{equation}
    \widetilde{S}_{\rm b}(\epsilon)= -h(\epsilon,N)\dfrac{(\epsilon-\epsilon_{\rm sh}(N))(\epsilon-\epsilon_{\rm sc}(N))}{\epsilon^2}
    \,,
    \label{eqn:general_ansatz_for_action}
\end{equation}
where $h(\epsilon,N)$ is expected to be a generic function with a value of order of one with a weak dependence on $N$. 

\begin{figure}
    \centering
    \includegraphics[width=\linewidth]{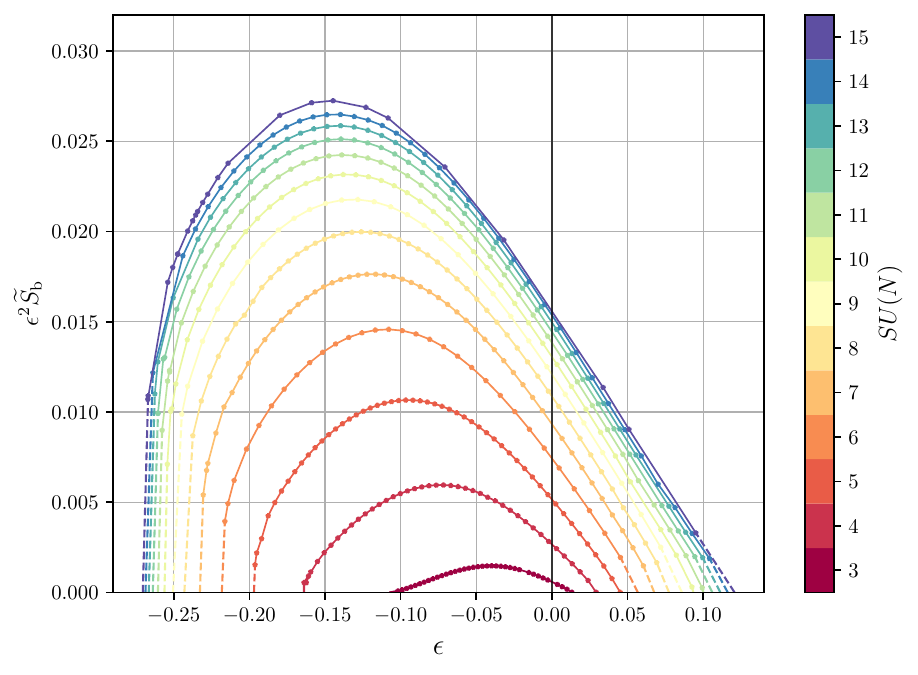}
    \caption{The rescaled bounce action $\epsilon^{2}\widetilde{S}_{\rm b}$ as a function of $\epsilon$ for $SU(N)$ with $N=\{3,4,\ldots,15\}$. $\widetilde{S}_{\rm b}$ has two zeros, one at $\epsilon_{\rm sc}$ and the other at $\epsilon_{\rm sh}$, and a double pole at $\epsilon=0$ as expected. The results are seen to converge at large $N$. Numerical values are calculated using the \texttt{FindBounce} package \cite{Guada:2020xnz}.}
    \label{fig:all_bounce}
\end{figure}
\begin{figure}
    \centering
    \includegraphics[width=\linewidth]{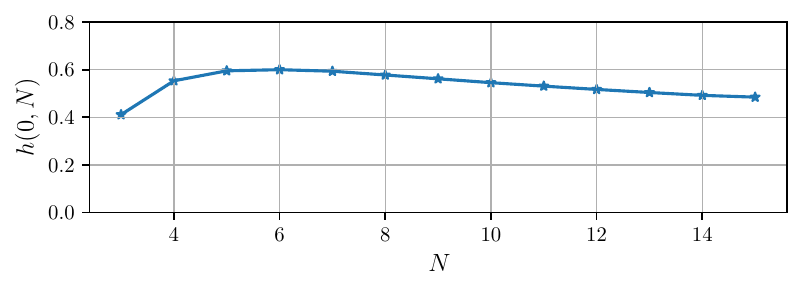}
    \caption{$h(0,N)$ as a function of $N$.}
    \label{fig:h_N_all}
\end{figure}

We study the bounce action for $SU(N)$ with $N$ ranging from 3 to 15 using the \texttt{FindBounce} package \cite{Guada:2020xnz}. The results are plotted in figure~\ref{fig:all_bounce}. Further, in the supplementary material, we also show an example of the bounce profile and trajectories for $N=10$ highlighting its non-trivial nature in the field space. We see from the figures that our expectation describes the PT in the softly broken SUSY case very well. We can use the form above at $\epsilon\to0$ to compare the action derived near $T_{\rm cr}$ from the lattice data as given in Eq.\eqref{eqn:o_3_ym}:
\begin{equation}
    \epsilon^{2}S_{\rm b}\Big|_{\epsilon\to 0}=-\dfrac{\Lambda^3 N^2}{\sigma_{\rm str}^2 L_{\rm cr}}h(0,N)\epsilon_{\rm sh}\epsilon_{\rm sc}\,.
    \label{eqn:thin_wall_sym}
\end{equation}
The values of $h(0,N)$ for different $N$ are given in figure~\ref{fig:h_N_all}, agreeing with our expectation that $h(0,N)\sim \mathcal{O}(1)$. Taking these lessons on the PT between centre-symmetric vs centre-broken phases in the softly broken $\mathcal{N}=1$ SYM on $\mathbb{R}^{3}\times S^{1}$, we can make an estimate of the supercooling bound in thermal YM theory.

\prlsec{Supercooling in Thermal YM} 
It has been argued  that the PT in the SYM theory can be continuously connected to the thermal confinement-deconfinement transition in YM theory~\cite{Poppitz:2012sw, Poppitz:2012nz,Anber:2013sga, Anber:2017tug}. Our expression in Eq.\eqref{eqn:thin_wall_sym} can be directly translated to thermal YM with $\sigma_{\rm str}$ identified as the string tension of the fundamental string in YM,
\begin{equation}
    \epsilon^{2}~S_{\rm b}^{\scym}\Big|_{\epsilon\to 0}\sim-\dfrac{N^2 T_{\rm cr}^4}{\sigma_{\rm str}^2}\epsilon_{\rm sh}^{\scym}\epsilon_{\rm sc}^{\scym}\,.
    \label{eq:YMactionfinal}
\end{equation}
Note that we use ``$\sim$" instead of ``$=$", as we cannot fix the value of the order of one number analogous to  $h(0,N)$ in the SYM case studied above. We also make the replacement $\Lambda \to T_{\rm cr}$, which is shown to be correct for any $N$~\cite{Allton:2008ty} on the lattice up to an order one number. This shows that the smallness of the lattice results in Eq.\eqref{eqn:o_3_ym} is connected to the smallness of the product $\epsilon_{\rm sh}\epsilon_{\rm sc}$. More precisely, using $T_{\rm cr}^2=0.36\sigma_{\rm str}$ \cite{Lucini:2005vg,Lucini:2012wq} gives:
\begin{equation}
    \epsilon_{\rm sh}^{\scym}\epsilon_{\rm sc}^{\scym}\sim -6.8\times 10^{-3}\,.
\end{equation}
To estimate the supercooling in the case of thermal YM, we note that in the softly broken SYM considered above, the magnitude of the $\epsilon$ for the superheating instability is larger than that for supercooling and only becomes comparable to it for a large value of $N$. Hence, a conservative approach would be to take $\vert\epsilon^{\scym}_{\rm sh}\vert\sim \epsilon^{\scym}_{\rm sc}$ in which case,
\begin{equation}
    \epsilon^{\scym}_{\rm sc}\sim 0.08\,.
\end{equation}
As mentioned above, this number is estimated up to $\mathcal{O}(1)$ factor, and hence pinning down the exact value using only analytical tools would be a challenging exercise. Fortunately, this prediction should be testable in lattice simulations and if true can have dramatic effects on the gravitational wave signatures from the phase transition.

\prlsec{Discussion} In this letter, we have proposed that the anomalously large confining-PT rate of $SU(N)$ Yang-Mills theory is due to a close-by instability of the metastable (de)confined phases. Based on hints from lattice data and calculable analogue models of confinement we have estimated the maximal supercooling achievable for the metastable deconfined phase for all $N\geq 3$. Our estimate shows that maximal supercooling is a few percent, hence we expect strong suppression of GW signals from such confinement PTs. In addition, two striking features of our prediction for the PT are the fast fall of $\epsilon^{2}S_{\rm b}$ near $\epsilon_{\rm sh}$ and the near-linear behaviour on the $\epsilon_{\rm sc}$ side as seen in figure \ref{fig:all_bounce}. It would be interesting if this behaviour can be tested on the lattice following on from the results of \cite{Seppa:2025lud}. From a cosmological perspective, our work is just an early step towards a full understanding of the confining PT dynamics, and thus of potential GW signals.  Important open questions include a better understanding of the thermodynamic and kinetic properties of the strongly coupled plasma that can have a significant effect on the dynamics, especially the specific heat of the metastable phases, thermal transport coefficients, and scattering rates on the bubble wall.
\\
\\
\textit{Acknowledgements} We thank Mike Teper and Kieran Twaites for continued discussions regarding many aspects of lattice results. GRK thanks Riikka Seppa for discussion of recent lattice studies of superheating. VL thanks Rashmish K. Mishra, Nicklas Ramberg, and other participants of the workshop ``Windows into New Physics in the Sky" for discussions.
PA and VL thank Munich Institute for Astro-, Particle and BioPhysics (MIAPbP) for support and hospitality funded by the Deutsche Forschungsgemeinschaft (DFG) Excellence Strategy – EXC-2094 – 390783311.
GRK expresses gratitude for support via the Somerville College Oxford Ryniker Lloyd Graduate Scholarship jointly with a Clarendon Fund Scholarship. The work of PA, VL, and JMR are supported by the STFC grant ST/X000761/1. For the purpose of Open Access, the authors have applied a CC BY public copyright licence to any Author Accepted Manuscript version arising from this submission.

\bibliographystyle{utphys}
\bibliography{references}
\clearpage
\onecolumngrid
\setcounter{equation}{0}
\setcounter{figure}{0}
\setcounter{table}{0}

\begin{center}
    {\Large \textbf{Supplemental Material for ``Prediction for Maximum Supercooling in SU(N) Confinement Transition''}}\\[0.5cm]
\end{center}

\section{Example of the Bounce Trajectory for $N=10$}
Here, we provide numerical details of the bounce trajectory for the \emph{deconfinement-confinement} phase transition (see the main text for the definition of these phases) for the illustrative case of $SU(10)$ softly-broken super Yang-Mills theory on $\mathbb{R}^{3}\times S^{1}$. 

\begin{figure}[h!]
    \centering
    \includegraphics[width=0.7\linewidth]{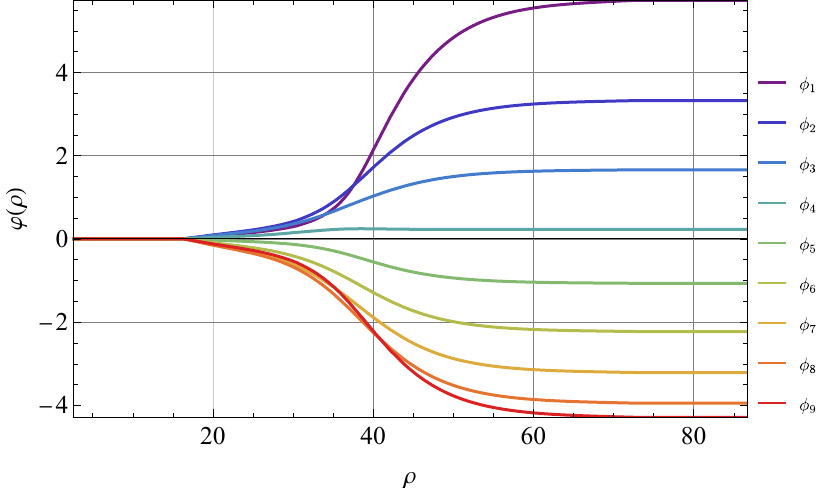}
    \caption{The bounce profile of nine holonomies $\vec{\phi}$ for $SU(10)$ at $\epsilon=0.037$ as function of $\rho$ that is the radial parameter of the $O(3)$-symmetric bounce.}
    \label{fig:bounce}
\end{figure}

In the controlled setup of the softly-broken supersymmetric theory, the phase transition is parametrized by the dynamical variables $\vec{\phi}$, the $N-1$ gauge holonomies (defined in equation (7) in the main text). We use the package \texttt{FindBounce}~\cite{Guada:2020xnz} to find the bounce profiles for the transition.
Figure \ref{fig:bounce} shows the profiles of these nine fields $\phi_{i}$, $i\in\{1,2,\cdots,9\}$ at an example value of $\epsilon=0.037$. The bounce has a \emph{thick wall} profile, which is typically true for all the cases we study. We do not use thin-wall approximation in any of our numerical results in the main text. 

\begin{figure}[h!]
    \centering
    \begin{subfigure}{0.45\linewidth}
        \centering
          \includegraphics[width=0.8\linewidth]{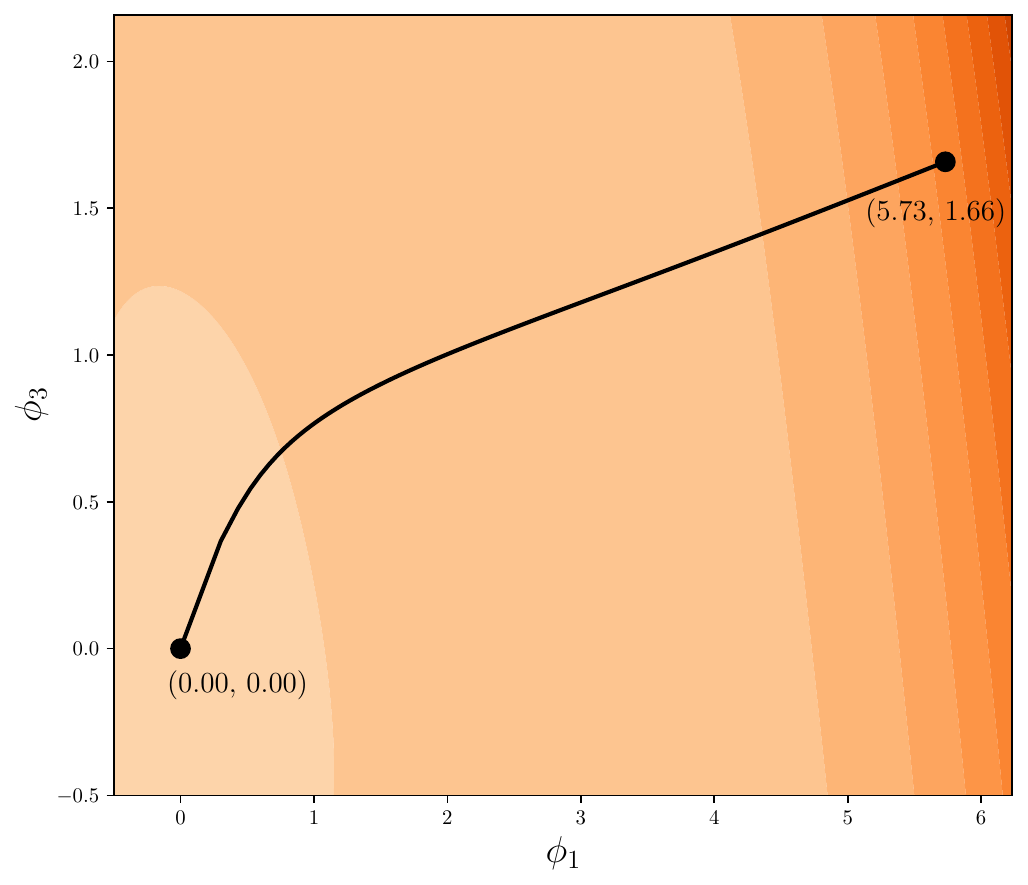}
    \end{subfigure}
    \begin{subfigure}{0.45\linewidth}
        \centering
          \includegraphics[width=0.8\linewidth]{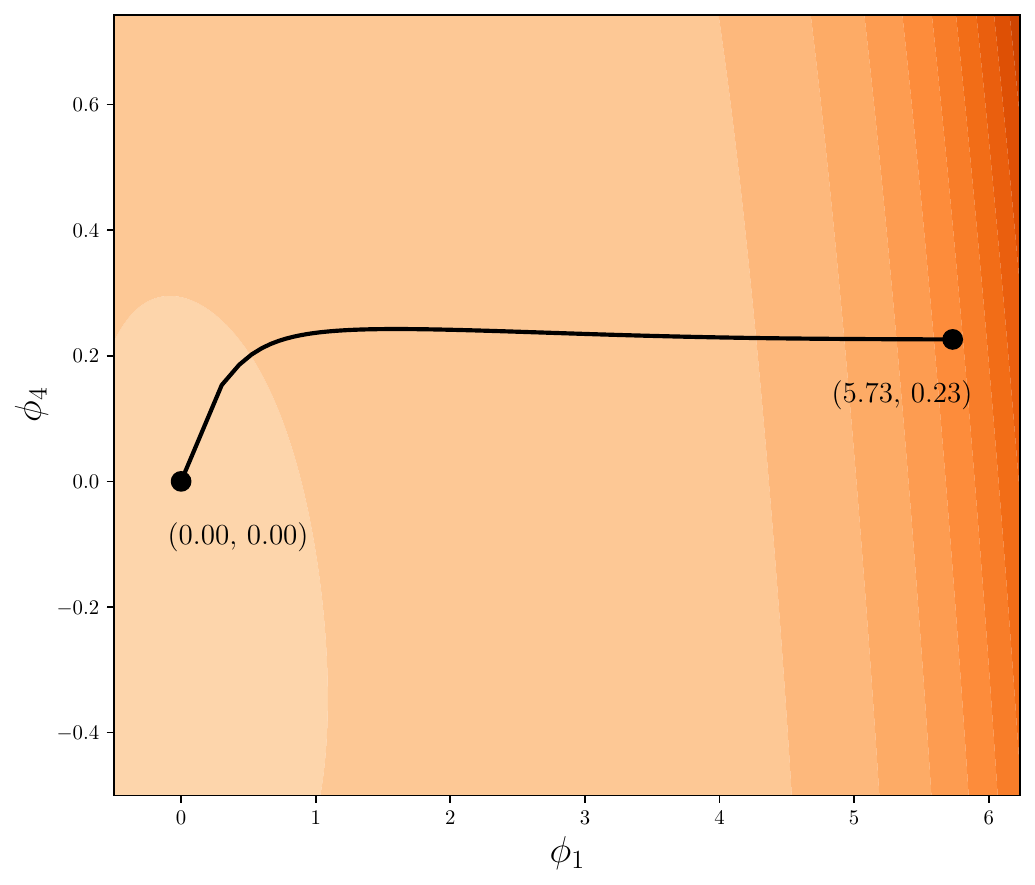}
    \end{subfigure}
    \begin{subfigure}{0.45\linewidth}
        \centering
          \includegraphics[width=0.8\linewidth]{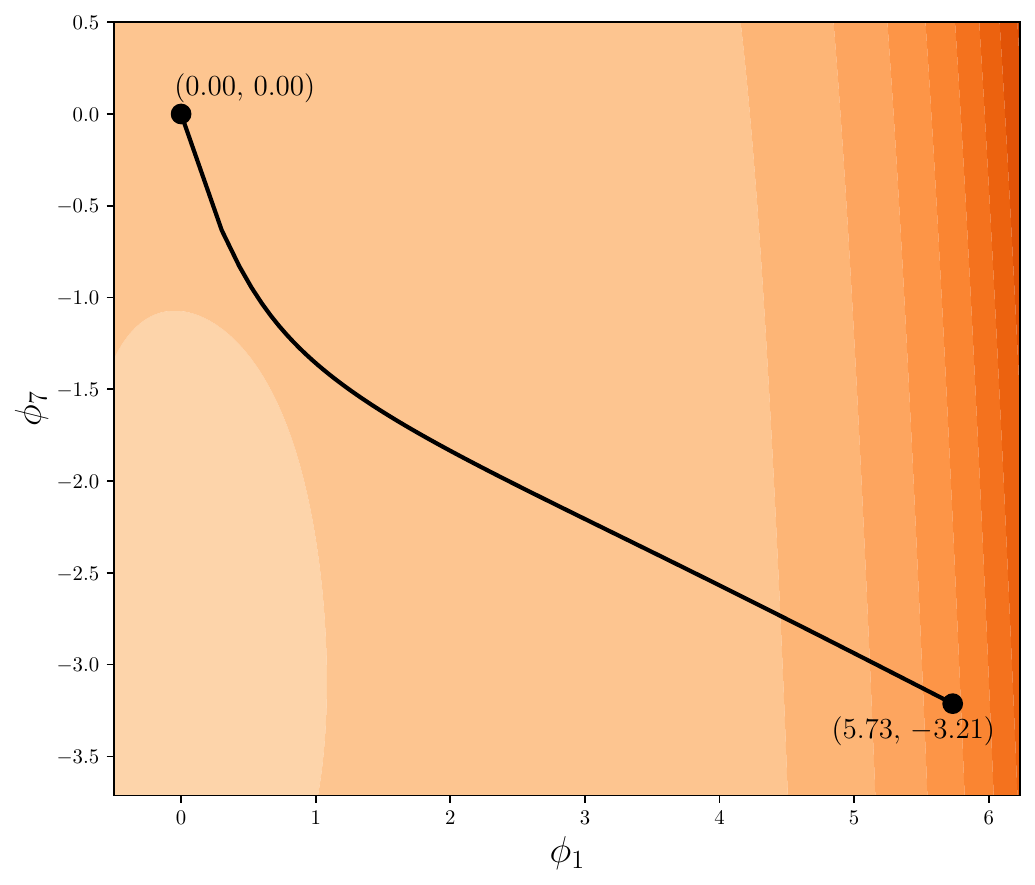}
    \end{subfigure}
    \begin{subfigure}{0.45\linewidth}
        \centering
          \includegraphics[width=0.8\linewidth]{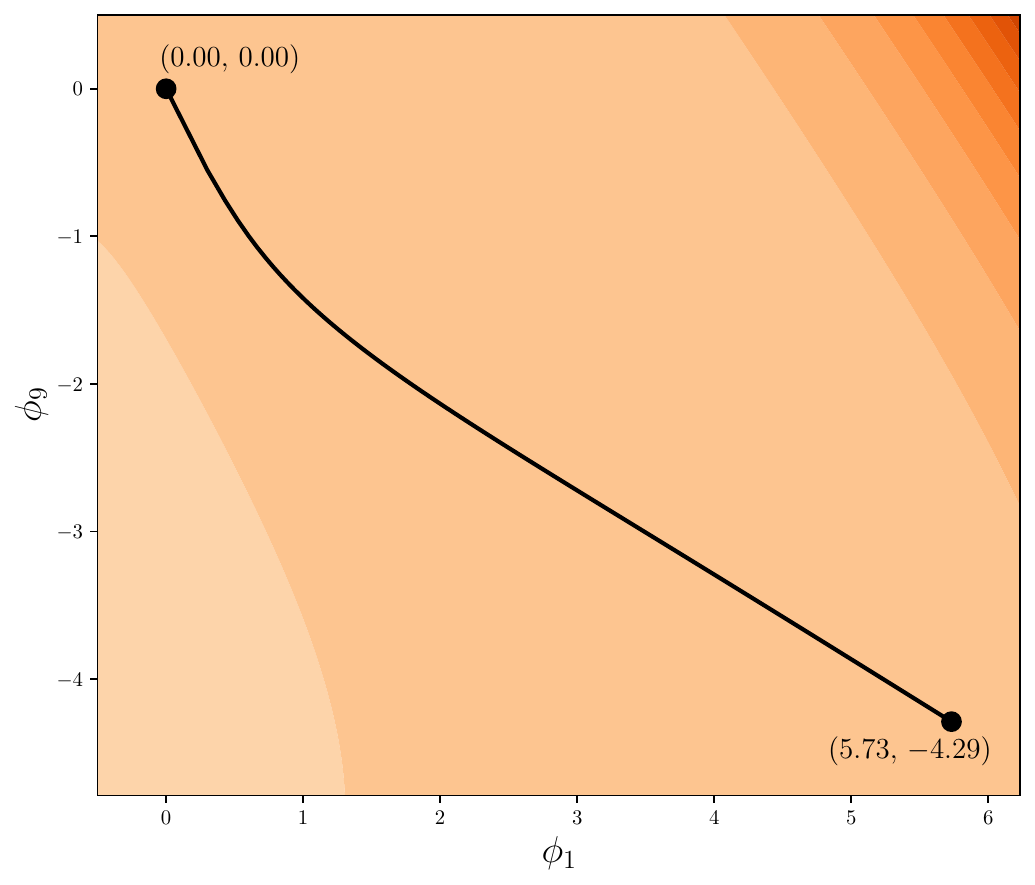}
    \end{subfigure}
    \caption{Projection of bounce trajectory for $SU(10)$ at $\epsilon=0.037$ on different planes. In the case of $SU(10)$, the field space has nine dimensions. \textbf{Upper Left}: $\phi_{1}-\phi_{3}$ plane, \textbf{Upper Right}: $\phi_{1}-\phi_{4}$ plane, \textbf{Lower Left}: $\phi_{1}-\phi_{7}$ plane, \textbf{Lower Right}: $\phi_{1}-\phi_{9}$ plane.}
    \label{fig:different_projections}
\end{figure}

In figure \ref{fig:different_projections} we show the projections of the bounce trajectory on different planes in the $\vec{\phi}$ field space. 
The bounce trajectory in the field space is not a straight line connecting the true and the false vacua. This is one of the crucial distinctions with studies of the thermal deconfinement-confinement transitions using Polyakov loop effective model (for example studies such as \cite{Huang:2020crf} and references within) where the potential is written is terms of a single field. Although the Polyakov loop is an order parameter, it is not the field controlling the dynamics of the transition. 
\vfill 
\pagebreak

\end{document}